\documentclass[twocolumn,preprintnumbers,amsmath,amssymb,floatfix,superscriptaddress,nofootinbib]{revtex4}

\usepackage{amssymb,amsmath,latexsym,
multirow,comment,appendix,slashed,color,afterpage,multirow,makecell}

\usepackage[colorlinks=true
,urlcolor=blue
,anchorcolor=blue
,citecolor=blue
,filecolor=blue
,linkcolor=blue
,menucolor=blue
,linktocpage=true
,pdfproducer=medialab
,pdfa=true
]{hyperref}

\pdfoutput=1
\usepackage{bm}
\usepackage{setspace}
\usepackage{tikz}
\usetikzlibrary{calc}

\usepackage{cleveref}
\crefname{table}{Table}{Tables}
\crefname{equation}{Eq.}{Eqs.}
\crefname{appendix}{App.}{Apps.}
\crefname{section}{Sec.}{Secs.}
\crefname{figure}{Fig.}{Figs.}
\usepackage[normalem]{ulem}

\newcommand{\Lag}{\mathcal{L}}
\newcommand{\Dcal}{\mathcal{D}}
\newcommand{\Amp}{\mathcal{A}}
\newcommand{\Mcal}{\mathcal{M}}
\newcommand{\Chris}{G}
\newcommand{\dd}{\text{d}}

\def\eg{\textit{e.g.}}
\def\ie{\textit{i.e.}}

\definecolor{colorTC}{rgb}{.2,.7,.2}

\newcommand{\s}{\hspace{0.8pt}}

\begin{document}

\vspace*{-30mm}

\title{On-Shell Covariance of Quantum Field Theory Amplitudes}

\author{Timothy Cohen}
\affiliation{Institute for Fundamental Science, University of Oregon, Eugene, OR 97403, USA \vspace{3pt}}

\author{Nathaniel Craig}
\affiliation{Department of Physics, University of California, Santa Barbara, CA 93106, USA\vspace{3pt}}

\author{Xiaochuan Lu}
\affiliation{Institute for Fundamental Science, University of Oregon, Eugene, OR 97403, USA \vspace{3pt}}

\author{Dave Sutherland}
\affiliation{INFN Sezione di Trieste, via Bonomea 265, 34136 Trieste TS, Italy \vspace{3pt}}
\affiliation{Theoretical Physics Department, CERN, 1211 Geneva 23, Switzerland}

\begin{abstract}
\begin{center}
{\bf Abstract}
\vspace{-7pt}
\end{center}
Scattering amplitudes in quantum field theory are independent of the field parameterization, which has a natural geometric interpretation as a form of `coordinate invariance.' Amplitudes can be expressed in terms of Riemannian curvature tensors, which makes the covariance of amplitudes under non-derivative field redefinitions manifest. We present a generalized geometric framework that extends this manifest covariance to {\it all} allowed field redefinitions. Amplitudes satisfy a recursion relation that closely resembles the application of covariant derivatives to increase the rank of a tensor. This allows us to argue that (tree-level) amplitudes possess a notion of `on-shell covariance,' in that they transform as a tensor under any allowed field redefinition up to a set of terms that vanish when the equations of motion and on-shell momentum constraints are imposed. We highlight a variety of immediate applications to effective field theories.
\end{abstract}

\maketitle

\begin{spacing}{1.1}

\noindent
It is well known that scattering amplitudes in quantum field theory are invariant under field redefinitions. The original demonstration of this statement dates back to the early 1960s, where the invariance of the $S$-matrix was shown for the case of `local' (no derivatives) and `almost local' (a finite number of derivatives) field redefinitions~\cite{Borchers1960, Chisholm:1961tha, Kamefuchi:1961sb, Epstein2008}.  The path integral provides an intuitive framing of this fact.  Since amplitudes are computed by performing a weighted integral over all possible field configurations, it is appropriate to think of the field that appears in the Lagrangian as an integration variable.  This implies that amplitudes should be independent of the way we choose to parameterize the field.  This simple picture is essentially correct, and was put on rigorous footing in~\cite{Arzt:1993gz}.

This invariance is often obscure at the Lagrangian level, but may be illuminated by geometry: reformulating field theories in terms of geometrically covariant objects can make the invariance of observables under field redefinitions manifest.  This so-called `(constant) field space geometry' has indeed been utilized in a variety of scenarios, as we will summarize below. However, this approach faces a major limitation in that it only accommodates field redefinitions without derivatives.

Our goal in this paper is to generalize the notion of field space geometry to incorporate the full set of allowed field redefinitions.  The key insight is to work with suitably defined off-shell quantities and to compute them in terms of functional derivatives.  The resulting `functional geometry' framework allows us to identify objects that are manifestly covariant under all allowed field redefinitions once the on-shell conditions are enforced.  In particular, suitably defined tree-level $S$-matrix elements are `on-shell covariant' in this sense.

Historically, there are three important implications of invariance under field redefinitions that played an essential role in the development of quantum field theory.  The first is the notion of gauge invariance, which is a special kind of field redefinition that leaves the Lagrangian explicitly invariant.  Indeed, arguing that observables are independent of the gauge choice for non-Abelian gauge theories was one of the main motivations of Ref.~\cite{Kamefuchi:1961sb}.  Field redefinitions are also extremely useful for the study of Effective Field Theory (EFT).  For example, the primary goal of Ref.~\cite{Coleman:1969sm} was to demonstrate that the linear and non-linear representations of the Goldstone fields were equivalent, which they achieved using field space geometry.  This led to the CCWZ formulation of EFT Lagrangians for the Goldstone boson sector of theories with spontaneously broken global symmetries~\cite{Callan:1969sn}.  Field redefinitions were also central to the development of more general EFTs.  One is free to use the equations of motion to write the EFT Lagrangian with the fewest number of derivative operators, which typically leads to a simpler form for the interactions~\cite{Georgi:1991ch}. The search for efficient ways to accommodate basis redundancies in EFTs continues to this day.

Geometry has been used in many guises to capture the effects of non-derivative field redefinitions; see \eg\ \cite{Burgess:2010zq, Alonso:2015fsp, Alonso:2016oah, Finn:2019aip}. It has seen something of a revival in recent years, largely due to Refs.~\cite{Alonso:2015fsp, Alonso:2016oah}, which introduced a geometric picture of the EFT for the scalar sector of the Standard Model of particle physics.  This so-called Higgs EFT (HEFT) provides a generic low energy approximation to extensions of the Standard Model that is expressed directly in the electroweak broken phase of the theory.  Ref.~\cite{Alonso:2015fsp} further showed how to express scattering amplitudes in terms of curvature tensors constructed from the field space metric of the HEFT Lagrangian. This picture has found various important applications, such as robustly classifying which HEFTs admit linear realizations of electroweak symmetry~\cite{Cohen:2020xca}, determining the scale of unitarity violation~\cite{Cohen:2021ucp, Alonso:2021rac} for the Standard Model scalar sector, providing results for low-point observables in the Standard Model Effective Field Theory to all orders in the fields~\cite{Helset:2020yio}, and explaining the soft theorems for generalized Goldstone bosons from a new point of view \cite{Cheung:2021yog}.

Yet the power of field space geometry to describe EFTs is limited by its inability to accommodate derivative field redefinitions. In this paper, we solve this problem by developing a generalized notion of field space geometry incorporating the full space of allowed field redefinitions. In particular, we construct a generalization of the field space metric and Christoffel symbol, which will be the essential building blocks for a novel expression of an off-shell recursion relation for computing amplitudes.  (The connection to the famous Berends-Giele off-shell recursion~\cite{Berends:1987me}, as well as the Vilkovisky-DeWitt effective action \cite{Vilkovisky:1984st, DeWitt:1985sg}, will be made below.)   This will allow us to make what we mean by `on-shell covariance' completely precise.  The derivations for the main results presented here, along with some first applications, will appear in a forthcoming companion paper~\cite{Paper2}.

\setcounter{section}{1}
\section{Covariant Amplitudes}
\label{sec:amplitude}
\noindent
We begin by defining scattering amplitudes in a way that will be useful for making their on-shell covariance manifest. The objects of interest are the amplitudes $\Amp$ that have been stripped of their LSZ residue factors:
\begin{equation}
\overline{\Mcal}(p_1, \cdots, p_n) \equiv -Z_\eta^{-n/2} \Amp(p_1, \cdots, p_n) \,.
\end{equation}\noindent
These objects are known to be covariant under field redefinitions \emph{without} derivatives, see \eg~\cite{Alonso:2015fsp, Alonso:2016oah, Cohen:2021ucp}. In this section, we will review the textbook methods for computing them directly from the path integral using so-called functional methods. This will allow us to show that such (tree-level) amplitudes are covariant under general field redefinitions, up to terms that vanish on shell, see~\cref{sec:geometry}.

We begin by defining the generating functional $W[J]$ for a generic theory of commuting fields $\eta(x)$, whose indices (such as those of flavor and spin) we suppress. In terms of a path integral,
\begin{equation}
e^{iW[J]} \equiv \int \Dcal\eta\, e^{i \left(S[\eta]+\int\! \dd^4x\s J(x)\eta(x) \right)} \,,
\label{eqn:EDef}
\end{equation}
where $\Dcal \eta$ is the functional integration measure, $S[\eta]$ is the action defined as a functional of $\eta$, and $J(x)$ is an external source. One can extract connected correlation functions of $\eta$ from this object by taking functional derivatives with respect to $J$.  For example, the vacuum expectation value is given by
\begin{equation}
\phi^x \equiv \langle\eta(x)\rangle_J = \frac{\delta W}{\delta J_x} \,,
\label{eqn:phixDef}
\end{equation}
where we have introduced the shorthand $J_x \equiv J(x)$ and $\phi^x \equiv \phi(x)$. The propagator is
\begin{equation}
iD^{xy} \equiv i\langle\eta(x)\eta(y)\rangle_{J,\,\text{conn}} = \frac{\delta^2 W}{\delta J_x\, \delta J_y} \,,
\label{eqn:DxyDef}
\end{equation}
where $\langle \dots \rangle_{J,\,\text{conn}}$ denotes that this is defined for general $J$, and only includes the connected contributions.

We will find it more convenient to work with the one-particle irreducible (1PI) effective action $\Gamma[\phi]$, which is the Legendre transform of $W[J]$:
\begin{equation}
-\Gamma \equiv J_x \phi^x - W \,,
\label{eqn:GammaDef}
\end{equation}
and hence satisfies
\begin{subequations}
\begin{align}
\frac{\delta(-\Gamma)}{\delta\phi^x} &= J_x \,, \label{eqn:Gamma1} \\[5pt]
\frac{\delta^2 (-\Gamma)}{\delta\phi^x \delta\phi^y} &= \left[ \frac{\delta^2 W}{\delta J_x \delta J_y} \right]^{-1} \equiv -iD_{xy}^{-1} \,, \label{eqn:Gamma2}
\end{align}
\end{subequations}
where $D^{xy}$ is the all-orders propagator. Note the summation convention $J_x \phi^x = \int \dd^4 x J(x) \phi(x)$.

To generate our objects of interest, we simply compute
\begin{equation}
\hspace{-6pt}\Mcal_{x_1 \cdots x_n}\! \equiv\! -\!\left(-iD_{x_1 y_1}^{-1}\right) \cdots \left(-iD_{x_n y_n}^{-1}\right) \frac{\delta^n W}{\delta J_{y_1} \cdots \delta J_{y_n}} \,,\!\!
\label{eqn:MGeneration}
\end{equation}
where $\Mcal_{x_1 \cdots x_n}$ are precisely the LSZ-residue-factor-stripped amplitudes in position space
\begin{align}
&(2\pi)^4 \delta^4\left(p_1+\cdots+p_n\right) \overline{\Mcal}(p_1, \cdots, p_n) \notag\\[5pt]
&\hspace{60pt} = \int \left[\s \prod_{i=1}^{n} \dd^4 x_i\, e^{ip_ix_i} \right] \Mcal_{x_1 \cdots x_n}\big|_{J=0} \,,
\label{eqn:AfromM}
\end{align}\noindent
where we have taken all external momenta to be outgoing. Note that the quantities $\Mcal_{x_1 \cdots x_n}$ defined in \cref{eqn:MGeneration} are functionals of $J$, and so we must impose an on-shell condition $J=0$ when going from $\Mcal \to \Amp$ (taking $J=0$ is equivalent to taking the origin defined in \cite{Coleman:1969sm}), which is equivalent to enforcing that $\phi$ satisfies the $\eta$ equations of motion through \cref{eqn:Gamma1}. Furthermore, the physical amplitudes $\Amp\left(p_1, \cdots, p_n\right)$ only correspond to observables for on-shell momenta $p_i$, while $\Mcal_{x_1 \cdots x_n}$ as defined in \cref{eqn:MGeneration} are off shell. This ambiguity is automatically removed via \cref{eqn:AfromM} by contracting with the one-particle wavefunctions $e^{i p_i x_i}$, which are zero eigenfunctions of $-i D^{-1}_{x_i y} \big|_{J=0}$ when the momenta $p_i$ are on shell.  This eliminates any terms proportional to an inverse propagator that acts on a leg in $\Mcal$. In summary, as far as the amplitudes $\Amp$ are concerned, only the `physical' pieces of $\Mcal_{x_1 \cdots x_n}$ contribute, which are obtained by enforcing the following two conditions
\begin{subequations}\label{eqn:OnShell}
\begin{align}
\frac{\delta(-\Gamma)}{\delta\phi^z} &= J_z = 0   \quad\text{(equations of motion)}, \\[4pt]
\frac{\delta^2(-\Gamma)}{\delta\phi^{x_i}\delta\phi^y}\bigg|_{J=0} &= -i D^{-1}_{x_i y}\bigg|_{J=0} = 0   \quad \text{(on-shell legs)}.
\end{align}
\end{subequations}
This defines what we mean by `on-shell' in this paper.

\section{Off-Shell Recursion}
\label{sec:recursion}
\noindent
The definitions presented in the previous section will naturally lead to a geometric interpretation.  To see this explicitly, we will use the fact that the $\Mcal_{x_1 \cdots x_n}$ satisfy the following tensor-like recursion relation (derived below):
\begin{align}
\Mcal_{x_1 \cdots x_n x} &= \frac{\delta}{\delta\phi^{x}} \Mcal_{x_1 \cdots x_n} - \sum_{i=1}^{n} \Chris_{x x_i}^y \Mcal_{x_1 \cdots \hat{x}_i y \cdots x_n} \notag\\[4pt]
& \equiv \nabla_x \Mcal_{x_1 \cdots x_n} \,,
\label{eqn:CovariantDerivative}
\end{align}
where the notation $x_1 \cdots \hat{x}_i y \cdots x_n$ denotes the string $x_1 \cdots x_n$ with $x_i$ replaced by $y$. As the second line in \cref{eqn:CovariantDerivative} implies, the right-hand side can be viewed as an analog of a covariant derivative on the field configuration space manifold, where
\begin{equation}
\Chris_{x_1 x_2}^y \equiv i D^{yz} \Mcal_{z x_1x_2} = iD^{yz}\, \frac{\delta^3 (-\Gamma)}{\delta\phi^z \delta\phi^{x_1} \delta\phi^{x_2}} \,.
\label{eqn:ChrisDef}
\end{equation}
We can interpret the covariant derivative in \cref{eqn:CovariantDerivative} as generating parallel transport on the field space manifold.  This motivates identifying $\Chris_{x_1 x_2}^y$ as the `functional Christoffel symbol.' In addition, the inverse propagator $-iD^{-1}_{xy}$ and the propagator $iD^{xy}$ can be viewed as the metric and inverse metric on this manifold.\footnote{As we will discuss in \cref{sec:geometry}, $\Chris_{x_1 x_2}^y$ and $-iD^{-1}_{xy}$ are indeed generalizations of the Christoffel symbol and metric on the constant field space manifold; see \cref{eqn:ChrisReduction,eqn:MetricReduction} below. Note that \cref{eqn:ChrisDef} has a factor of two relative to the definition of a metric connection.}

We emphasize that the recursion relation \cref{eqn:CovariantDerivative} holds off-shell, in particular for general $J\ne0$, which means that the functional Christoffel symbol $\Chris_{x x_i}^y$ encodes all the interactions in the theory, not just the cubic ones. It can be algebraically derived from \cref{eqn:MGeneration} by repeatedly applying
\begin{equation}
\left[ \frac{\delta}{\delta J_y} \;,\; iD^{y_iz_i} \right] = - \left(iD^{yw}\right) \left(iD^{y_iw_i}\right) G_{ww_i}^{z_i} \,,
\end{equation}
see~\cite{Paper2} for details. Alternatively, it can also be obtained diagrammatically, as we will now show.  An amplitude $\Mcal_{x_1 \cdots x_n}$ can be constructed by gluing together the following two types of ingredients
\begin{subequations}
\begin{align}
k\text{-point 1PI vertices}:   &\qquad   -i\frac{\delta^k (-\Gamma)}{\delta\phi^{y_1} \cdots \delta\phi^{y_k}} \,, \\[5pt]
\text{(full) propagators}:     &\qquad   D^{y_1y_2}  \,.
\end{align}
\end{subequations}
Consider adding an additional leg to an amplitude, labeled by $x$.  There are multiple ways this can be accomplished.  One can connect this leg to a particular 1PI vertex
\newcommand{\rad}{0.3}
\newcommand{\dotrad}{0.43}
\newcommand{\dotradM}{0.4}
\newcommand{\bigrad}{0.9}
\newcommand{\leglen}{0.2}
\tikzset{vertexstyle/.style={circle,draw,thick,inner sep=2pt}}
\tikzset{dotstyle/.style={thick,dotted}}
\begin{align}
  \mathord{
  \begin{tikzpicture}[baseline=-0.65ex]
  \node[vertexstyle] (centralnode) at (0,0) {\footnotesize 1PI};
  \node[vertexstyle,gray] (node1) at (180:1.2) {\footnotesize 1PI};
  \node[vertexstyle,gray] (node2) at (135:1.2) {\footnotesize 1PI};
  \draw (centralnode) -- node[pos=0.2,shift={(270:0.2)}] {\scriptsize $y_1$} (node1);
  \draw (centralnode) -- node[pos=0.2,shift={(45:0.2)}] {\scriptsize $y_m$} (node2);
  \draw[dotstyle,gray] ([shift=(90:\dotrad)]node1) arc (90:270:\dotrad);
  \draw[dotstyle,gray] ([shift=(45:\dotrad)]node2) arc (45:225:\dotrad);
  \draw[dotstyle] ([shift=(140:\dotrad)]centralnode) arc (140:175:\dotrad);
  \draw (centralnode) -- (70:0.7) node[shift={(70:0.2)}] {\scriptsize $y_{m+1}$};
  \draw (centralnode) -- (25:0.7) node[shift={(25:0.2)}] {\scriptsize $y_{k}$};
  \draw[dotstyle] ([shift=(30:\dotrad)]centralnode) arc (30:65:\dotrad);
\end{tikzpicture}} \hspace*{-3ex}
  \quad\longrightarrow\quad &
  \mathord{
  \begin{tikzpicture}[baseline=-0.65ex]
  \node[vertexstyle] (centralnode) at (0,0) {\footnotesize 1PI};
  \node[vertexstyle,gray] (node1) at (180:1.2) {\footnotesize 1PI};
  \node[vertexstyle,gray] (node2) at (135:1.2) {\footnotesize 1PI};
  \draw (centralnode) -- node[pos=0.2,shift={(270:0.2)}] {\scriptsize $y_1$} (node1);
  \draw (centralnode) -- node[pos=0.2,shift={(45:0.2)}] {\scriptsize $y_m$} (node2);
  \draw[dotstyle,gray] ([shift=(90:\dotrad)]node1) arc (90:270:\dotrad);
  \draw[dotstyle,gray] ([shift=(45:\dotrad)]node2) arc (45:225:\dotrad);
  \draw[dotstyle] ([shift=(140:\dotrad)]centralnode) arc (140:175:\dotrad);
  \draw (centralnode) -- (70:0.7) node[shift={(70:0.2)}] {\scriptsize $y_{m+1}$};
  \draw (centralnode) -- (25:0.7) node[shift={(25:0.2)}] {\scriptsize $y_{k}$};
  \draw[dotstyle] ([shift=(30:\dotrad)]centralnode) arc (30:65:\dotrad);
  \draw (centralnode) -- (0:0.7) node[shift={(0:0.2)}] {\scriptsize $x$};
  \end{tikzpicture}} \, ,
\end{align}
which is the diagrammatic representation of
\begin{align}
-i\frac{\delta^k (-\Gamma)}{\delta\phi^{y_1} \cdots \delta\phi^{y_k}} \quad\longrightarrow\quad &-i\frac{\delta^{k+1} (-\Gamma)}{\delta\phi^{y_1} \cdots \delta\phi^{y_k}\delta\phi^x} \notag\\[5pt]
&= \frac{\delta}{\delta\phi^x} \left[ -i\frac{\delta^k (-\Gamma)}{\delta\phi^{y_1} \cdots \delta\phi^{y_k}} \right] \,.
\end{align}
Another option is to split a propagator into two propagators using an insertion of the three-point 1PI vertex:
\begin{align}
  \mathord{
  \begin{tikzpicture}[baseline=-0.65ex]
  \node[vertexstyle] (node1) at (-0.8,0) {\footnotesize 1PI};
  \node[vertexstyle] (node2) at (0.8,0) {\footnotesize 1PI};
  \draw (node1) -- node[pos=0.2,shift={(270:0.2)}] {\scriptsize $y_1$} node[pos=0.8,shift={(270:0.2)}] {\scriptsize $y_2$} (node2);
  \draw[dotstyle] ([shift=(90:\dotrad)]node1) arc (90:270:\dotrad);
  \draw[dotstyle] ([shift=(-90:\dotrad)]node2) arc (-90:90:\dotrad);
\end{tikzpicture}}
\quad\longrightarrow\quad &
\mathord{
\begin{tikzpicture}[baseline=-0.65ex]
\node[vertexstyle] (node1) at (-1.4,-0.2) {\footnotesize 1PI};
\node[vertexstyle] (node2) at (1.4,-0.2) {\footnotesize 1PI};
\node[vertexstyle] (centralnode) at (0,0.2) {\footnotesize 1PI};
\draw (node1) -- node[pos=0.2,shift={(285:0.2)}] {\scriptsize $y_1$} node[pos=0.8,shift={(285:0.2)}] {\scriptsize $z_1$} (centralnode);
\draw (node2) -- node[pos=0.2,shift={(255:0.2)}] {\scriptsize $y_2$} node[pos=0.8,shift={(255:0.2)}] {\scriptsize $z_2$} (centralnode);
\draw[dotstyle] ([shift=(90:\dotrad)]node1) arc (90:270:\dotrad);
\draw[dotstyle] ([shift=(-90:\dotrad)]node2) arc (-90:90:\dotrad);
\draw (centralnode) -- (90:0.9) node[shift={(90:0.2)}] {\scriptsize $x$};
\end{tikzpicture}}
\end{align}
which is the same as
\begin{align}
D^{y_1y_2} \quad\longrightarrow\quad &
D^{y_1z_1} \left[ -i\frac{\delta^3(-\Gamma)}{\delta\phi^{z_1}\delta\phi^x\delta\phi^{z_2}} \right] D^{z_2y_2} \notag \\[5pt]
&= \frac{\delta}{\delta\phi^x} D^{y_1y_2} \,.
\end{align}
Putting together all the possible ways of performing both types of these insertions amounts to taking the functional derivative $\frac{\delta}{\delta\phi^x}$ of the amplitude, namely the first term in \cref{eqn:CovariantDerivative}.

However, these two operations do not cover all the ways of adding a leg to an amplitude. One can also attach a three-point 1PI vertex to a leg, which turns that leg into a propagator that is connected with two new legs
\begin{align}
  \mathord{
  \begin{tikzpicture}[baseline=-0.65ex]
  \node[vertexstyle,fill=gray] (centralnode) at (0,0) {\footnotesize $\Mcal$};
  \draw (centralnode) -- (180:0.7) node[shift={(180:0.2)}] {\scriptsize $x_{i}$};
  \draw (centralnode) -- (45:0.7) node[shift={(45:0.2)}] {\scriptsize $x_{1}$};
  \draw (centralnode) -- (-45:0.7) node[shift={(-45:0.2)}] {\scriptsize $x_{n}$};
  \draw[dotstyle] ([shift=(50:\dotradM)]centralnode) arc (50:175:\dotradM);
  \draw[dotstyle] ([shift=(185:\dotradM)]centralnode) arc (185:310:\dotradM);
  \end{tikzpicture}}
  \quad\longrightarrow\quad &
  \mathord{
  \begin{tikzpicture}[baseline=-0.65ex]
  \node[vertexstyle,fill=gray] (centralnode) at (0,0) {\footnotesize $\Mcal$};
  \node[vertexstyle] (newnode) at (-1.2,0) {\footnotesize 1PI};
  \draw (centralnode) -- node[pos=0.2,shift={(270:0.2)}] {\scriptsize $y$} node[pos=0.8,shift={(270:0.17)}] {\scriptsize $z$} (newnode);
  \draw (centralnode) -- (45:0.7) node[shift={(45:0.2)}] {\scriptsize $x_{1}$};
  \draw (centralnode) -- (-45:0.7) node[shift={(-45:0.2)}] {\scriptsize $x_{n}$};
  \draw (newnode) -- +(135:0.7) node[shift={(135:0.2)}] {\scriptsize $x$};
  \draw (newnode) -- +(-135:0.7) node[shift={(-135:0.2)}] {\scriptsize $x_{i}$};
  \draw[dotstyle] ([shift=(50:\dotradM)]centralnode) arc (50:175:\dotradM);
  \draw[dotstyle] ([shift=(235:\dotradM)]centralnode) arc (235:310:\dotradM);
  \end{tikzpicture}}
  \, ,
\end{align}
or in terms of expressions
\begin{align}
  \Mcal_{x_1 \cdots x_i \cdots x_n} \quad\longrightarrow\quad & -i\frac{\delta^3(-\Gamma)}{\delta\phi^{x}\delta\phi^{x_i}\delta\phi^{z}} D^{zy} \Mcal_{x_1 \cdots \hat{x}_i y \cdots x_n} \notag\\[5pt]
  & = -G^y_{x x_i} \Mcal_{x_1 \cdots \hat{x}_i y \cdots x_n} \,.
\end{align}
The functional Christoffel symbol terms in \cref{eqn:CovariantDerivative} are accounting for all possible ways of doing this operation. A more detailed derivation of this recursion relation will be presented in~\cite{Paper2}.

It is worth noting that our recursion relation \cref{eqn:CovariantDerivative} reduces to the Berends-Giele recursion relation~\cite{Berends:1987me, Brown:1992ay, Monteiro:2011pc} when $J=0$ is enforced. To make this connection more concrete, we extend \cref{eqn:ChrisDef} to define an analog of the generalized Christoffel symbols
\begin{equation}
\Chris_{x_1 \cdots x_n}^y \equiv i D^{yz} \Mcal_{z x_1\cdots x_n} \,,
\label{eqn:GenChrisDef}
\end{equation}
which, similar to $\Mcal$, satisfy the following recursion relation for general $J\ne0$
\begin{align}
\Chris^z_{x_1 \cdots x_n x} &= \frac{\delta}{\delta\phi^{x}} \Chris^z_{x_1 \cdots x_n} - \sum_{i=1}^{n} \Chris_{x x_i}^y \Chris^z_{x_1 \cdots \hat{x}_i y \cdots x_n} \notag\\[4pt]
& \equiv \nabla_x \Chris^z_{x_1 \cdots x_n} \,,
\label{eqn:CovariantDerivativeGenChris}
\end{align}
as expected for the usual generalized Christoffel symbols.

We see from \cref{eqn:MGeneration,eqn:AfromM} that as far as the physical amplitudes $\Amp\left(p_1, \cdots, p_n\right)$ are concerned, it is sufficient to study these generalized Christoffel symbols at $J=0$. On the other hand, through \cref{eqn:phixDef,eqn:MGeneration,eqn:GenChrisDef}, one can derive that the quantities $\Chris_{x_1 \cdots x_n}^y \vert_{J=0}$ play a central role in the relation between $\phi^y$ and the ``raised'' source field $\hat J^x \equiv \left( i D^{xy} \vert_{J=0} \right) J_y$:
\begin{equation}
\phi^y = \hat J^y - \sum_{n=2}^\infty \frac{1}{n!} \left( \Chris_{x_1 \cdots x_n}^y \vert_{J=0} \right) \hat J^{x_1} \cdots \hat J^{x_n} \,.
\end{equation}
Therefore, one can derive $\Chris_{x_1 \cdots x_n}^y \vert_{J=0}$ (and hence obtain the physical amplitudes $\Amp\left(p_1, \cdots, p_n\right)$) by computing $\phi^y[\hat J^y]$ order-by-order in $\hat J^y$. Doing this at the tree-level by iteratively solving the equation of motion in \cref{eqn:Gamma1} about $J=0$ is the Berends-Giele recursion procedure \cite{Berends:1987me, Brown:1992ay, Monteiro:2011pc}. This approach efficiently computes the quantities $\Chris_{x_1 \cdots x_n}^y \vert_{J=0}$, and hence the physical amplitudes. Our \cref{eqn:CovariantDerivativeGenChris} (or equivalently \cref{eqn:CovariantDerivative}) reveals that $\Chris_{x_1 \cdots x_n}^y$ possess a more general recursion structure that holds also in the case $J\ne0$. This is the insight that allows us to define the generalized notion of field space geometry that can accommodate derivative field redefinitions.

\section{Functional Geometry}
\label{sec:geometry}
\noindent
The tensor-like recursion relation for the quantities $\Mcal_{x_1 \cdots x_n}$ in \cref{eqn:CovariantDerivative} implies the existence of a geometric structure on the field configuration space manifold, which we call `functional geometry.' The objects $\Mcal_{x_1 \cdots x_n}$ indeed transform as tensors, up to terms that vanish on shell, in the sense of \cref{eqn:OnShell}. We now have everything we need to make this statement precise.

We can parameterize a general field redefinition $\phi(x) \to \tilde{\phi}(x)$ that could include derivatives using a functional relation $\phi[\tilde\phi]$.  If we restrict to tree-level amplitudes, the 1PI effective action transforms as a scalar
\begin{equation}
\tilde\Gamma[\tilde\phi] = \tilde{S}[\tilde\phi] = S\big[\phi[\tilde\phi]\big] = \Gamma\big[\phi[\tilde\phi]\big] \,.
\label{eqn:treeActionScalar}
\end{equation}
This allows us to derive the following transformation law for the amplitudes~\cite{Paper2}:
\begin{equation}
\widetilde{\Mcal}_{x_1 \cdots x_n} = \left( \frac{\delta\phi^{y_1}}{\delta\tilde\phi^{x_1}} \cdots \frac{\delta\phi^{y_n}}{\delta\tilde\phi^{x_n}} \right) \Mcal_{y_1 \cdots y_n} + U_{x_1 \cdots x_n} \,,
\label{eqn:OnShellCovar}
\end{equation}
where
\begin{align}
U_{x_1 \cdots x_n} &= a_{x_1 \cdots x_n y_1} \frac{\delta(-\Gamma)}{\delta\phi^{y_1}} \nonumber\\[4pt]
&\hspace{20pt}+ \sum_{i=1}^n b_{x_1 \cdots \hat{x}_i \cdots x_n y_1} \frac{\delta\phi^{y_2}}{\delta\tilde\phi^{x_i}} \frac{\delta^2(-\Gamma)}{\delta\phi^{y_1}\delta\phi^{y_2}} \,,
\label{eqn:U}
\end{align}
for some $a$ and $b$, are a set of `evanescent' terms that vanishes if the field redefinition $\phi[\tilde\phi]$ preserves the two on-shell conditions in \cref{eqn:OnShell}. In other words, on-shell covariance implies that $\Mcal$ transforms like a tensor up to a set of evanescent terms that do not contribute to physical observables. In fact, $\Chris_{x_1 \dots x_n}^y$ and $-iD_{x_1x_2}^{-1}$ only transform as a Christoffel symbol and a metric in the same sense, \ie,\ up to evanescent terms.

Finally, we will show how this generalization of field space geometry reduces to the conventional objects. If we restrict to the case of only allowing for field redefinitions without derivatives, then the quantity $\Chris_{x_1 x_2}^y$ defined in \cref{eqn:ChrisDef} indeed reduces to the Christoffel symbol in the constant field space geometric picture explored in Refs.~\cite{Alonso:2015fsp, Alonso:2016oah, Finn:2019aip}.
To see this, we use the generic form of the Lagrangian
\begin{equation}
\Lag = - V + \frac12\s g_{ab}\! \left(\partial_\mu\phi^a\right) \left(\partial^\mu\phi^b\right) \,,
\label{eqn:ScalarLag}
\end{equation}
to derive $\Chris_{x_1 x_2}^y$ explicitly. Making flavor indices ($a,b,\ldots$) explicit, it satisfies
\begin{align}
& \lim_{q^2\to\infty} \int \dd^4x_1\, \dd^4x_2\, \dd^4y\, e^{ip_1x_1+ip_2x_2} e^{-iqy} \notag\\
&\hspace{80pt} \times \left[ G_{ab}^c\left(x_1, x_2, y\right)\big|_{\partial_\mu\phi_i=0} \right] \notag\\[5pt]
&\hspace{20pt} = (2\pi)^4 \delta^4(p_1+p_2-q)\,\notag\\
&\hspace{80pt} \times \frac12\, g^{cd} \left( g_{da,b} + g_{db,a} - g_{ab,d} \right) \,.
\label{eqn:ChrisReduction}
\end{align}
This tells us that $\Chris_{x_1 x_2}^y$ is a generalization of the Christoffel symbol used in Refs. \cite{Alonso:2015fsp, Alonso:2016oah, Finn:2019aip}. Similarly, the inverse propagator $-iD_{x_1x_2}^{-1}$ is a generalization of the metric on the constant field space manifold
\begin{align}
& \int \dd^4x_1\, \dd^4x_2\, e^{ip_1x_1} e^{-ip_2x_2} \left[-iD_{ab}^{-1}(x_1, x_2) \bigg|_{\partial_\mu\phi_i=0} \right]
\notag\\[7pt]
&\hspace{20pt} = (2\pi)^4 \delta^4(p_1-p_2)\, \left( -p_2^2\, g_{ab} + V_{,ab} \right) \,.
\label{eqn:MetricReduction}
\end{align}
This shows that we have indeed generalized the building blocks of Riemannian geometry as promised.

\section{Conclusions}
\label{sec:conclusions}
\noindent
In this paper, we have presented a generalization of field space geometry that accommodates field redefinitions involving derivatives.  The key building blocks are the functional metric and functional Christoffel symbol introduced here.  These objects enabled us to write down a covariant derivative on the field space manifold.  When acting on an amplitude, the parallel transport generated by this covariant derivative yields a new amplitude with an additional leg.  This provides a new type of off-shell recursion relation for computing amplitudes.  We then leveraged this recursion relation to demonstrate that tree-level amplitudes have a manifest notion of on-shell covariance, namely that they transform like tensors up to terms which vanish when the on-shell conditions are enforced.  This gives us a new way of understanding the invariance of amplitudes under field redefinitions reframed in terms of functional geometry.

There are many settings where this generalization can be applied.  One immediate application is the generalization of invariant criteria for EFTs to linearly realize a symmetry \cite{Alonso:2015fsp, Alonso:2016oah, Cohen:2021ucp} to accommodate derivative field redefinitions.  We claim that if there exists a fixed point $\phi_0(x)$ on the field configuration space manifold, which corresponds to the point where the symmetry can be linearly realized, then we can solve for it using
\begin{equation}
\frac{\delta^n(-\Gamma)}{\delta\pi_{x_1}\cdots\delta\pi_{x_n}} \bigg|_{\phi=\phi_0(x)} = 0 \,,
\end{equation}
where the $\pi$'s are the Goldstone directions on the manifold. The existence of the fixed point $\phi_0(x)$ determined using this criterion can not be obscured using derivative field redefinitions anywhere on the field space manifold.  Our framework should similarly generalize our understanding of how Lagrangian terms map onto kinematic structures in amplitudes, and allow the generalization of the analysis of soft-theorems presented in Ref.~\cite{Cheung:2021yog} to accommodate the full set of allowed field redefinitions.

This paper opens many directions for future research.  Given the supporting details and explicit examples in \cite{Paper2}, one obvious extension will be to understand how the functional geometry manifests beyond tree-level.  We know that such a framework must exist, since it is well known that amplitudes are invariant under field redefinitions to any order in perturbation theory.  Although our recursion relation \cref{eqn:CovariantDerivative} holds to all orders in perturbation theory, the 1PI effective action is no longer invariant beyond tree level, \ie, \cref{eqn:treeActionScalar} does not hold.  An understanding of the covariance properties of the loop level 1PI effective action will likely make it clear how to geometrize the loop level amplitudes as well.

Another area that is worth studying is the connection to the Vilkovisky-DeWitt formulation of the path integral \cite{Vilkovisky:1984st, DeWitt:1985sg}.  Their approach is to make the covariance under field redefinitions (without derivatives) manifest off-shell by modifying the coupling between the field and the source that appears in the path integral, see \cref{eqn:EDef}.  It would be very interesting to understand how to incorporate the new methods provided by functional geometry into this way of writing the path integral to make covariance under all allowed field redefinitions manifest. Aspects of our construction are also reminiscent of DeWitt's manifestly covariant formulation of quantum field theory \cite{DeWitt:1984sjp, DeWitt:2003pm} which we will explore further in \cite{Paper2}.

Functional geometry provides a language with which to understand the on-shell covariance of amplitudes in quantum field theory.  Given the foundational role played by the freedom to perform field redefinitions when defining our theories, we are optimistic that the new approach proposed here will lead to further insights.

\subsection*{Note Added}

As we were finishing this paper, we were made aware of the exciting related work by Cheung, Helset, and Parra-Martinez \cite{Cheung:2022xxx}.  These authors propose the existence of a `geometry-kinematics' duality, which allows them to identify the building blocks of a generalized geometry that accommodates field redefinitions including derivatives.  Our constructions differ at the detailed level.  For example, as the authors note, the geometric invariants that define their kinematic geometry depend on choices made when specifying the initial metric, which is not the case for the functional geometry developed here.  It will be very interesting to understand the relation between these two approaches.  We are grateful to the authors for sharing their draft with us.

\acknowledgments
We are grateful to Zvi Bern, Cliff Burgess, and Aneesh Manohar for useful comments on this draft.
The work of T.~Cohen and X.~Lu is supported by the U.S.~Department of Energy under grant number DE-SC0011640.
The work of N.~Craig is supported by the U.S.~Department of Energy under the grant DE-SC0011702.
D.~Sutherland has received funding from the European Union's Horizon 2020 research and innovation programme under the Marie Skłodowska-Curie grant agreement No.~754496. He thanks CERN for hospitality.
N.~Craig thanks LBNL and the BCTP for hospitality at the beginning of this work.

\end{spacing}

\bibliographystyle{utphys}
\bibliography{ref}

\end{document}